\documentclass[aip,reprint]{revtex4-1}
\usepackage{graphicx}% Include figure files
\usepackage{dcolumn}% Align table columns on decimal point
\usepackage{bm}% bold math
\usepackage{amsmath}

\draft % marks overfull lines with a black rule on the right

\begin{document}
%\preprint{}

\title{Neutral gas temperature maps of the pin-to-plate Argon micro discharge into the ambient air.} %Title of paper



\author{S. F. Xu}
\affiliation{The State Key Laboratory on Fiber Optic Local Area, Communication Networks and Advanced Optical Communication Systems, Key Laboratory for Laser Plasmas and Department of Physics and Astronomy, Shanghai Jiao Tong University, Shanghai 200240, China}
%\thanks{}
\author{X. X. Zhong}
\email{xxzhong@sjtu.edu.cn}
\affiliation{The State Key Laboratory on Fiber Optic Local Area, Communication Networks and Advanced Optical Communication Systems, Key Laboratory for Laser Plasmas and Department of Physics and Astronomy, Shanghai Jiao Tong University, Shanghai 200240, China}

\author{Asif Majeed}
\affiliation{The State Key Laboratory on Fiber Optic Local Area, Communication Networks and Advanced Optical Communication Systems, Key Laboratory for Laser Plasmas and Department of Physics and Astronomy, Shanghai Jiao Tong University, Shanghai 200240, China}
\affiliation{Department of Physics, University of Azad Jammu and Kashmir Muzaffarabad (A. K), Pakistan.}
%\collaboration{Department of Physics and Astronomy, Key Laboratory for Laser Plasmas (Ministry of Education), The State Key Laboratory on Fiber Optic Local Area, Communication Networks and Advanced Optical Communication Systems, Shanghai Jiao Tong University - Shanghai 200240, PRC}

\date{\today}

\begin{abstract}
This study is designed to explore the two dimensional temperature maps of the atmospheric argon discharge consisting of pin-to-plane electrodes supplied by a high voltage DC source. After checking the stability of the micro discharge, the two dimensional image plane focused by a quartz lens was scanned by the fiber probe driven by a 3D Mobile Platform. The rotational and vibrational temperatures are calculated using nitrogen emissions collected by the high resolution spectrometer and high sensitive intensified charge coupled device (ICCD). The rotational temperature varies from 1558.15 K to 2621.14 K and vibrational temperature varies from 3010.38 K to 3774.69 K, indicating a great temperature gradient due to small discharge size. The temperature maps show a lateral expansion and a sharp truncation in the radial direction. A double layers discharge is identified, where an arc discharge coats the glow discharge.
\end{abstract}

\pacs{52.70.-m, 52.30.-q, 52.80.Hc, 52.80.Mg}% plasma diagnostics, plasma flows, glow discharge, arc discharge. insert suggested PACS numbers in braces on next line

\maketitle %\maketitle must follow title, authors, abstract and \pacs

\section{Introduction}
Atmospheric pressure plasmas have received much attention as a promising source in numerous applications recently\cite{Li2006,Shi2008,Li2011,Ostrikov2005,Ostrikov2011,Ostrikov2013,Bivscan2015}. They can be used as anatytical tools, photonic devices, in the surface treatment of vulnerable materials, bacterial inactivation processes, surface modification, chemical vapor deposition, plasma etching on films\cite{Pearton1994,Sankaran2001,Lukins2001,Wu2007,Kyung2007,Deng2014}.

Characteristics of micro plasmas, such as neutral gas temperature, electron temperature, electron density, ion density, electric field and so on, play an important role in physical and chemical processes\cite{Petrov2002,Qi2009,Qi2014}. There are many studies concerning the relationship between the bulk gas temperature and control parameters, for example, discharge current and gas flux, where the gas temperature is estimated using radiative transitions of diatomic molecules\cite{Faure1998,Cruden2002,Staack2005,Bai2006,Lu2013,Xu2013}. However, the temperature mentioned above is the mean temperature without spatial resolution. Papers seldom reported the spatial resolution temperature maps of micro plasmas (micro discharges) although which can give more information of micro discharge mechanism and help optimization of micro discharge device for applications.

In this work, an experiment was designed to obtain the two dimensional gas temperature maps of micro plasmas generated in the atmosphere of the ambient air between a miniature flow argon and a metal film. Firstly voltage-current curve was measured. Then the stability of the micro discharge was checked and image plane focused by a quartz lens was scanned by a fiber probe and two dimensional spectrums of nitrogen emissions were got by a high resolution spectrometer and an intensified charge coupled device (ICCD). At last we got the two dimensional temperature maps of micro discharges suggesting a double layers coated discharge.
%#####################
\section{Experiment}
\begin{figure}
\center
\includegraphics{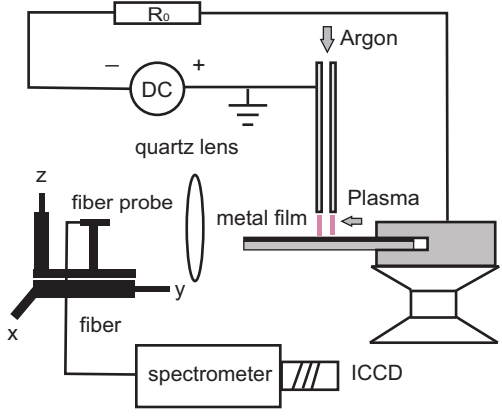}
\caption{Schematic diagram of the experimental device, not scaled.}
\label{device}
\center
\end{figure}

The schematic of the experimental device is shown in Fig.~\ref{device}. Argon gas flow jets into the ambient air through a stainless-steel capillary acting as the anode, whose internal diameter is 175 $\mu m$ and external diameter is one sixteenth of an inch. A metal film with smooth rounded edges serves as the cathode. The gap distance from the exit of capillary to the metal film is 2 $mm$. The circuit is driven by a high-voltage DC source with a ballast resistor limiting the discharge current. The argon gas flux is set as 10 $sccm$ controlled by a mass flow controller. So the mean speed of gas jet at the exit nearly equals 6.93 $m/s$. The titanium film used in this work is achieved by low temperature deposition on silicon substrate using reactive DC magnetron sputtering. More details about the deposition equipments can be found in the referance\cite{Zhou2007}.
%################
%%%%%%%%%%%%%%%%%

The primary methodology to get two dimensional temperature maps is that fiber probe driven by a 3D Mobile Platform scans the image plane focused by a quartz lens. Because signals of spontaneous radiation collected by the fiber probe is very weak. A high sensitive and enhancement ICCD (intensified charge coupled device, PI-MAX 4) is used to read the spectrum data made by a high resolution spectrometer (2400 groove/mm grating).
%%%%%%%%%%%%%%%%%%%%%%%
\begin{figure}
\center
\includegraphics{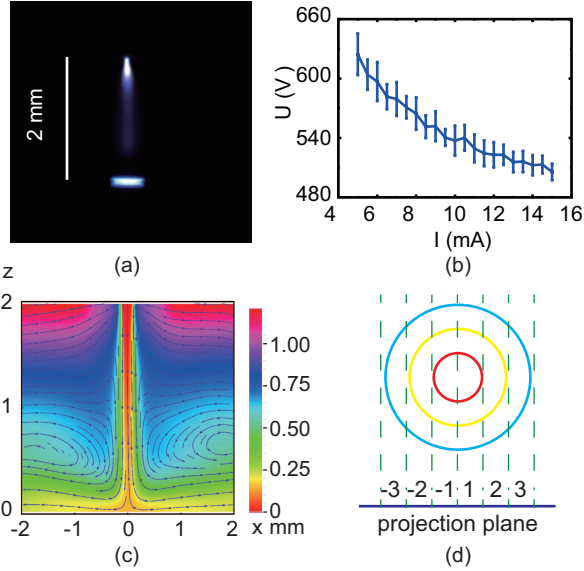}
\caption{(a) Discharge image taken by the Manta G201C CCD with the commercial lenses. (b) The voltage-current curves of micro discharges. The error bars are the standard deviation of six measurements. (c) Air density distribution obtained by solving full component Naiver-Stokes equations. (d) the sketch map of radiation projection.}
\label{fig2}
\center
\end{figure}
%%%%%%%%%%%%%%%%%%%%%%%

The typical DC discharge image is taken by a Manta G201C CCD with commercial image lenses at the discharge current of 10 $mA$ shown in Fig.~\ref{fig2} (a). The visible discharge structure is similar to that of a typical DC glow discharge, namely a negative glow, a Faraday dark space and a positive column are observed. Furthermore, Fig.~\ref{fig2} (b) gives the voltage-current curve, which is the average of six measurements. The error bars are the standard deviations of six measurements. The discharge voltage drops slowly as the discharge increases suggesting a characteristic of arc discharge different from the glow discharge. The discharge image and voltage-current curve indicate the complex nature of the argon discharge.

In this study, the second positive systems of nitrogen were used to estimate the rotational and vibrational temperatures assuming Boltzmann distribution of rotational and vibrational states population. We had better learn knowledge of the flow-field and air spatial distribution before discharge. The full component Naiver-Stokes equations were solved to estimate the air or nitrogen density distribution. The governing conservation equations are written below in cartesian coordinates:
\begin{equation}
\frac{\partial U}{\partial t}+\frac{\partial E}{\partial x}+\frac{\partial F}{\partial z}=J
\end{equation}
where vectors $U$, $E$ and $F$ are:
\begin{equation}
\begin{split}
&U=\begin{bmatrix}\rho\\ \rho u\\ \rho v\\ \rho E_t\\ \rho c_1\end{bmatrix},\\
&E=\begin{bmatrix}\rho u\\ \rho u^2+p-\tau_{xx}\\ \rho uv-\tau_{xz}\\ [\rho E_t+p-\tau_{xx}]u-v\tau_{xz}-k\frac{\partial T}{\partial x}-\rho D\triangle h\frac{\partial c_1}{\partial x}\\ \rho uc_1-\rho D\frac{\partial c_1}{\partial x}\end{bmatrix},\\
&F=\begin{bmatrix}\rho v\\ \rho uv-\tau_{zx}\\ \rho v^2+p-\tau_{zz}\\ [\rho E_t+p-\tau_{zz}]v-u\tau_{zx}-k\frac{\partial T}{\partial z}-\rho D\triangle h\frac{\partial c_1}{\partial z}\\ \rho vc_1-\rho D\frac{\partial c_1}{\partial z}\end{bmatrix},\\
&J=\begin{bmatrix}0\\ \rho f_x\\ \rho f_z\\ \rho(f_xu+f_zv)\\ 0\end{bmatrix}.
\end{split}
\label{controlequation}
\end{equation}
and $\rho$ is the mixture gas density, $u$, $v$ are the $x$ and $z$ components of velocity respectively and $V=\sqrt{u^2+v^2}$ is the speed, $E_t=e+\frac{V^2}{2}$ is the mixture gas energy density, $\tau_{xx}$ and $\tau_{zz}$ are the mixture gas normal viscous stresses, $\tau_{xz}$ and $\tau_{zx}$ are the mixture gas shear stresses. $\triangle h=(c_{p1}-c_{p2})T$ is the enthalpy difference between argon and air, $c_{p1}$ and $c_{p2}$ are specific heat at constant pressure of argon and air respectively. D is the binary diffusion coefficient, $c_1$ and $c_2$ are argon and air mass fraction, $c_2=1-c_1$. $f_x$ and $f_z$ are the $x$ and $z$ components of volume force.

Fig.~\ref{fig2} (c) shows the air density distribution. The air occupied the whole space until turning on the argon supply. The argon gas spread outwards from the exit and brought an argon channel while the air was put away. However, the argon took over the argon channel rather than the whole space. The more further away the argon channel, the more air or nitrogen. Fig.~\ref{fig2} (c) just shows the prominent aspect of the air distribution in two dimensional. In fact, jet flow and discharge are three dimensional. Even x-z plane spatially resolved, the light on the image plane is the projective signals in the y direction. The projection geometry is shown in Fig.~\ref{fig2} (d).

\begin{figure}
\center
\includegraphics{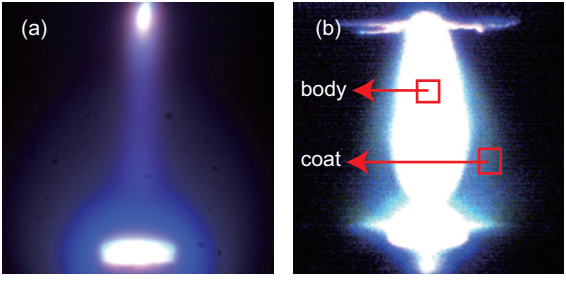}
\caption{Saturated exposure discharge images. (a) discharge image taken by the Manta G201C CCD with the quartz lens. (b) discharge image taken by the Manta G201C CCD with the commercial lenses.}
\label{fig3}
\center
\end{figure}

After knowing the projective nature of radiation, the spatial resolution of two dimensional spectrum depends on the quality of picture on image plane focused by the quartz lens. Fig.~\ref{fig3} (a) shows the discharge image on image plane directly taken by the Manta G201C CCD with a high gain number and long exposure time, that is to say, a saturated exposure image. Different from Fig.~\ref{fig2} (a), an obvious halo was found and the faraday dark space became unclear. At same time, a saturated exposure image was also taken by the Manta G201C CCD with the commercial lenses shown by the Fig.~\ref{fig3} (b). A coated discharge was identified where the discharge coat (coat discharge) surrounds the discharge body (body discharge). The nature of the coat discharge and the body discharge will be discussed in the future.

Because two dimensional spectrums were collected non-simultaneously, the stability of spectrum should be checked. The continuous spectrum from 392 nm to 400 nm was considered, which is emitted by $N_2(C^3\Pi_u,v'=2\to B^3\Pi_g,v''=5)$ and $N_2(C^3\Pi_u,v'=1\to B^3\Pi_g,v''=4)$ where $C^3\Pi_u,\ B^3\Pi_g$ denote electron configuration, $v',\ v''$ denote vibrational states. Emission intensity of rotational and vibrational bands is proportional to spontaneous emission coefficient and population of up state,
\begin{equation}
I_{Bv''J''}^{Cv'J'}=N_{v'J'}\frac{hc}{\lambda_{v'J'v''J''}}A_{v'J'v''J''},
\end{equation}
where $A_{v'J'v''J''}$ is spontaneous emission coefficient, $N_{v'J'}$ is population of up state, $\lambda_{v'J'v''J''}$ is wavelength, $v'J'$ denote the upper vibrational and rotational state, $v''J''$ denote the lower vibrational and rotational state, $h$ is the planck constant, $c$ is the speed of light.
\begin{figure}
\center
\includegraphics{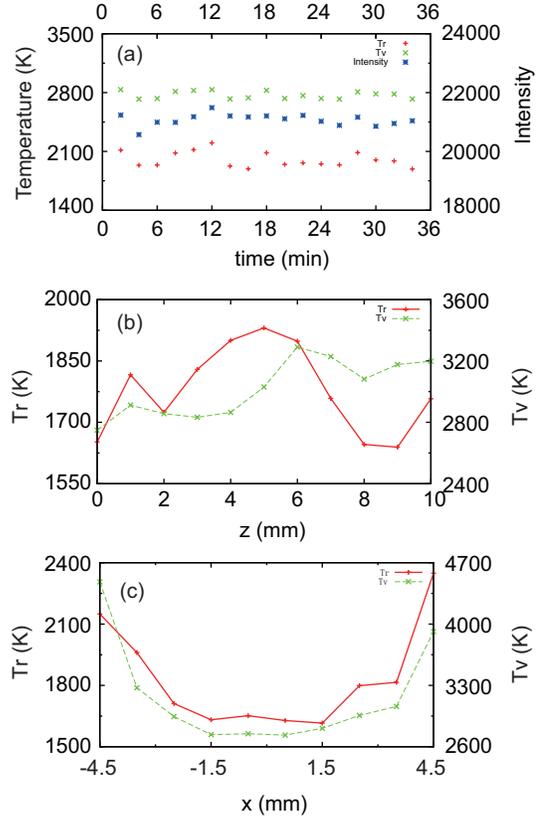}
\caption{(a) curves of rotational, vibrational temperature and counts number versus time. (b) rotational and vibrational temperature distribution in the z direction. (c) lateral rotational and vibrational temperature distributions at $z=0$.}
\label{fig4}
\center
\end{figure}
Under the Born-Oppenheimer approximation and Boltzmann distribution assumption, the intensity can be written as
\begin{equation}
I_{Bv''J''}^{Cv'J'}=\frac{D}{\lambda^{4}}q_{v',v''}exp(-\frac{E_{v'}}{kT_{v}})S_{J',J''}exp(-\frac{E_{J'}}{kT_{r}}),
\label{In}
\end{equation}
where $D$ is the constant, $q_{v',v''}$ is the Franck-Condon factor.  $S_{J',J''}$ is the line strength intensity. $E_{v'}$ and  $E_{J'}$ are vibrational state energy and rotational state energy. The more details can be found in references\cite{Phillips1976,Hartmann1978,Bai2006,Xu2013}.

Fig.~\ref{fig4} (a) shows curves of the rotational temperature, vibrational temperature and the intensity versus time of the test point on image plane. The intensity is the radiation intensity from 392 nm to 400 nm and recovered from the spectral data:
\begin{equation}
Intensity(y,z)=\frac{Counts(y,z)}{time\times gain}
\end{equation}
where $time$ is the integration time and $gain$ is the amplify gain.

The averages of rotational and vibrational temperatures are 2003.08 and 2770.32 K respectively. And the ratios between the standard deviation to the average of rotational temperature and vibrational temperature are $4.60\%$ and $1.64\%$ respectively. The rotational temperature and vibrational temperature do not depend on time, so the micro discharge can be regarded as stable.

Fig.~\ref{fig4} (b) shows the one dimensional temperature distribution along the axis of the steel capillary in the z direction. The rotational temperature and vibrational temperature decrease in the Faraday dark space, grow up in the positive column region and then decreases again. Fig.~\ref{fig4} (c) shows the lateral one dimensional temperature distribution at the $z=0$. It is found that the outer temperature is higher than inter temperature.

\begin{figure}
\center
\includegraphics{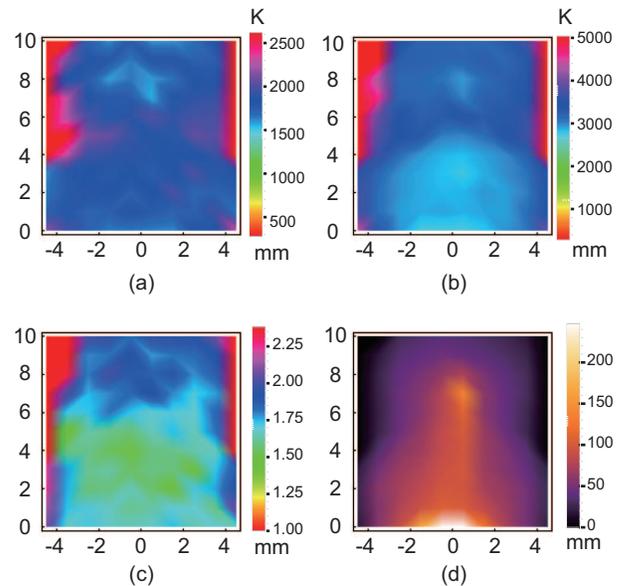}
\caption{(a) Two dimensional map of the rotational temperature. (b) Two dimensional map of the vibrational temperature. (c) Two dimensional map of ratio between the rotational temperature and the vibrational temperature. (d) The square root of discharge intensity on the wave band from 392 to 400 $nm$ recovered from spectral data. }
\label{map}
\center
\end{figure}

Fig.~\ref{map} (a) and Fig.~\ref{map} (b) show the two dimensional distribution of rotational temperature and vibrational temperature, respectively. The rotational temperature varies from 1558.15 K to 2621.14 K and vibrational temperature varies from 3010.38 K to 3774.69 K. So a great temperature gradient exits in the small size discharge, which maybe produce large heat exchange. In addition, Fig.~\ref{map} (c) shows the ratio map between rotational and vibrational temperature suggesting the disequilibrium degree of discharge. Fig.~\ref{map} (d) shows the discharge image on wave band from 392 nm to 400 nm recovered from spectral data in square root scale. The most significant features of temperature maps are the sharp truncation and the lateral expansion. The sharp truncation of temperature maps indicate a sudden interface between micro discharge and ambient air. In other words, there is not a continuous transition region connecting the ambient air and the micro discharge.

The Fig.~\ref{map} (d) shows that intensity decreases rapidly in the radial direction, but the signals are still detected by the ICCD in the outer region resulting from the coat discharge. So the lateral expansions are observed in temperature maps. The outer temperature is higher than inter temperature, as noticed, the voltage-current curve suggested an arc discharge. Therefore, we can infer that the body discharge is the glow discharge while the coat discharge is arc discharge.

The air or nitrogen density distribution shown in Fig.~\ref{fig2} (c) brings a question why nitrogen emission image looks like Fig.~\ref{map} (d). According to Eq.~\ref{In}, under the measured temperature distribution, the outer intensity should be stronger than inner intensity (obviously shown by Fig.~\ref{fig3} (c)), as opposed to Fig.~\ref{map} (d). How does nitrogen disappear or look like less. A more possible reason is that little nitrogen molecules participate in the coat discharge. The coat discharge may be in filamentary mode unlike the dispersive body discharge. Further more, the main excitation sources of nitrogen are electron, excited argon atom and ion. Besides electron excitation process, the heat transport through collision between nitrogen molecules and excited argon atoms may play an more important role. As we see from the temperature maps, the temperature is not maximum at the same x location, but the argon density and intensity are maximum. The energy transport of plasma particles is very complex, so the effect of collision excitation processes should be explored by detailed simulation.
%%%%%%%%%%%%%%%%%%%%%%%%%%%%%%%%%%%%%%%%
%%%%%%%%%%%%%%%%%%%%%%%%%%%%%%%%%%%%%%
%##############subsection
%\subsection{Internal energy increment calculation}
%###############section
\section{Conclusion}
In conclusion, experiments were designed to get the two dimensional spectrums of the micro atmospheric argon discharge on image plane collected by the high resolution spectrometer and high sensitive ICCD. And then the two dimensional temperature maps of the micro plasma are obtained. The rotational temperature varies from 1558.15 K to 2621.14 K and vibrational temperature varies from 3010.38 K to 3774.69 K. So a great temperature gradient exits in the small size discharge, which will produce large heat exchange. The temperature maps show a lateral expansion and a sharp truncation in the radial direction. The temperature maps and saturated exposure discharge images suggest that a filamentary arc coat discharge surround the body glow discharge. And the double layers coated discharge is identified.

The saturated exposure discharge images show a clear edge between the body and coat discharge, which is not reflected by the temperature maps. More higher spatial resolution researches are waited to study. And the two dimensional electron temperature maps is being prepared.
\begin{acknowledgments}
XXZ and SFX acknowledge support from the NSFC (Grant No. 11275127, 90923005), STCSM (Grant No. 09ZR1414600) and MOST of China.
\end{acknowledgments}

%\bibliography{Tmap}

\begin{thebibliography}{25}%
\makeatletter
\providecommand \@ifxundefined [1]{%
 \@ifx{#1\undefined}
}%
\providecommand \@ifnum [1]{%
 \ifnum #1\expandafter \@firstoftwo
 \else \expandafter \@secondoftwo
 \fi
}%
\providecommand \@ifx [1]{%
 \ifx #1\expandafter \@firstoftwo
 \else \expandafter \@secondoftwo
 \fi
}%
\providecommand \natexlab [1]{#1}%
\providecommand \enquote  [1]{``#1''}%
\providecommand \bibnamefont  [1]{#1}%
\providecommand \bibfnamefont [1]{#1}%
\providecommand \citenamefont [1]{#1}%
\providecommand \href@noop [0]{\@secondoftwo}%
\providecommand \href [0]{\begingroup \@sanitize@url \@href}%
\providecommand \@href[1]{\@@startlink{#1}\@@href}%
\providecommand \@@href[1]{\endgroup#1\@@endlink}%
\providecommand \@sanitize@url [0]{\catcode `\\12\catcode `\$12\catcode
  `\&12\catcode `\#12\catcode `\^12\catcode `\_12\catcode `\%12\relax}%
\providecommand \@@startlink[1]{}%
\providecommand \@@endlink[0]{}%
\providecommand \url  [0]{\begingroup\@sanitize@url \@url }%
\providecommand \@url [1]{\endgroup\@href {#1}{\urlprefix }}%
\providecommand \urlprefix  [0]{URL }%
\providecommand \Eprint [0]{\href }%
\providecommand \doibase [0]{http://dx.doi.org/}%
\providecommand \selectlanguage [0]{\@gobble}%
\providecommand \bibinfo  [0]{\@secondoftwo}%
\providecommand \bibfield  [0]{\@secondoftwo}%
\providecommand \translation [1]{[#1]}%
\providecommand \BibitemOpen [0]{}%
\providecommand \bibitemStop [0]{}%
\providecommand \bibitemNoStop [0]{.\EOS\space}%
\providecommand \EOS [0]{\spacefactor3000\relax}%
\providecommand \BibitemShut  [1]{\csname bibitem#1\endcsname}%
\let\auto@bib@innerbib\@empty
%</preamble>
\bibitem [{\citenamefont {Li}\ \emph {et~al.}(2006)\citenamefont {Li},
  \citenamefont {Lim}, \citenamefont {Kang},\ and\ \citenamefont
  {Uhm}}]{Li2006}%
  \BibitemOpen
  \bibfield  {author} {\bibinfo {author} {\bibfnamefont {S.-Z.}\ \bibnamefont
  {Li}}, \bibinfo {author} {\bibfnamefont {J.-P.}\ \bibnamefont {Lim}},
  \bibinfo {author} {\bibfnamefont {J.~G.}\ \bibnamefont {Kang}}, \ and\
  \bibinfo {author} {\bibfnamefont {H.~S.}\ \bibnamefont {Uhm}},\ }\href
  {\doibase http://dx.doi.org/10.1063/1.2355428} {\bibfield  {journal}
  {\bibinfo  {journal} {Physics of Plasmas (1994-present)}\ }\textbf {\bibinfo
  {volume} {13}},\ \bibinfo {eid} {093503} (\bibinfo {year}
  {2006})}\BibitemShut {NoStop}%
\bibitem [{\citenamefont {Shi}, \citenamefont {Wang},\ and\ \citenamefont
  {Wang}(2008)}]{Shi2008}%
  \BibitemOpen
  \bibfield  {author} {\bibinfo {author} {\bibfnamefont {H.}~\bibnamefont
  {Shi}}, \bibinfo {author} {\bibfnamefont {Y.}~\bibnamefont {Wang}}, \ and\
  \bibinfo {author} {\bibfnamefont {D.}~\bibnamefont {Wang}},\ }\href {\doibase
  http://dx.doi.org/10.1063/1.3033754} {\bibfield  {journal} {\bibinfo
  {journal} {Physics of Plasmas (1994-present)}\ }\textbf {\bibinfo {volume}
  {15}},\ \bibinfo {eid} {122306} (\bibinfo {year} {2008})}\BibitemShut
  {NoStop}%
\bibitem [{\citenamefont {Li}\ \emph {et~al.}(2011)\citenamefont {Li},
  \citenamefont {Wu}, \citenamefont {Yan}, \citenamefont {Wang},\ and\
  \citenamefont {Uhm}}]{Li2011}%
  \BibitemOpen
  \bibfield  {author} {\bibinfo {author} {\bibfnamefont {S.-Z.}\ \bibnamefont
  {Li}}, \bibinfo {author} {\bibfnamefont {Q.}~\bibnamefont {Wu}}, \bibinfo
  {author} {\bibfnamefont {W.}~\bibnamefont {Yan}}, \bibinfo {author}
  {\bibfnamefont {D.}~\bibnamefont {Wang}}, \ and\ \bibinfo {author}
  {\bibfnamefont {H.~S.}\ \bibnamefont {Uhm}},\ }\href {\doibase
  http://dx.doi.org/10.1063/1.3643224} {\bibfield  {journal} {\bibinfo
  {journal} {Physics of Plasmas (1994-present)}\ }\textbf {\bibinfo {volume}
  {18}},\ \bibinfo {eid} {103502} (\bibinfo {year} {2011})}\BibitemShut
  {NoStop}%
\bibitem [{\citenamefont {Ostrikov}(2005)}]{Ostrikov2005}%
  \BibitemOpen
  \bibfield  {author} {\bibinfo {author} {\bibfnamefont {K.}~\bibnamefont
  {Ostrikov}},\ }\href {\doibase 10.1103/RevModPhys.77.489} {\bibfield
  {journal} {\bibinfo  {journal} {Rev. Mod. Phys.}\ }\textbf {\bibinfo {volume}
  {77}},\ \bibinfo {pages} {489--511} (\bibinfo {year} {2005})}\BibitemShut
  {NoStop}%
\bibitem [{\citenamefont {Ostrikov}, \citenamefont {Cvelbar},\ and\
  \citenamefont {Murphy}(2011)}]{Ostrikov2011}%
  \BibitemOpen
  \bibfield  {author} {\bibinfo {author} {\bibfnamefont {K.~K.}\ \bibnamefont
  {Ostrikov}}, \bibinfo {author} {\bibfnamefont {U.}~\bibnamefont {Cvelbar}}, \
  and\ \bibinfo {author} {\bibfnamefont {A.~B.}\ \bibnamefont {Murphy}},\
  }\href@noop {} {\bibfield  {journal} {\bibinfo  {journal} {Journal of Physics
  D: Applied Physics}\ }\textbf {\bibinfo {volume} {44}},\ \bibinfo {pages}
  {174001} (\bibinfo {year} {2011})}\BibitemShut {NoStop}%
\bibitem [{\citenamefont {Ostrikov}, \citenamefont {Neyts},\ and\ \citenamefont
  {Meyyappan}(2013)}]{Ostrikov2013}%
  \BibitemOpen
  \bibfield  {author} {\bibinfo {author} {\bibfnamefont {K.}~\bibnamefont
  {Ostrikov}}, \bibinfo {author} {\bibfnamefont {E.~C.}\ \bibnamefont {Neyts}},
  \ and\ \bibinfo {author} {\bibfnamefont {M.}~\bibnamefont {Meyyappan}},\
  }\href {\doibase 10.1080/00018732.2013.808047} {\bibfield  {journal}
  {\bibinfo  {journal} {Advances in Physics}\ }\textbf {\bibinfo {volume}
  {62}},\ \bibinfo {pages} {113--224} (\bibinfo {year} {2013})}\BibitemShut
  {NoStop}%
\bibitem [{\citenamefont {Bi{\v{s}}{\'c}an}\ \emph {et~al.}(2015)\citenamefont
  {Bi{\v{s}}{\'c}an}, \citenamefont {Kregar}, \citenamefont {Cvelbar},
  \citenamefont {Mozeti{\v{c}}}, \citenamefont {Milo{\v{s}}evi{\'c}} \emph
  {et~al.}}]{Bivscan2015}%
  \BibitemOpen
  \bibfield  {author} {\bibinfo {author} {\bibfnamefont {M.}~\bibnamefont
  {Bi{\v{s}}{\'c}an}}, \bibinfo {author} {\bibfnamefont {Z.}~\bibnamefont
  {Kregar}}, \bibinfo {author} {\bibfnamefont {U.}~\bibnamefont {Cvelbar}},
  \bibinfo {author} {\bibfnamefont {M.}~\bibnamefont {Mozeti{\v{c}}}}, \bibinfo
  {author} {\bibfnamefont {S.}~\bibnamefont {Milo{\v{s}}evi{\'c}}},  \emph
  {et~al.},\ }\href@noop {} {\bibfield  {journal} {\bibinfo  {journal}
  {Spectrochimica Acta Part B: Atomic Spectroscopy}\ }\textbf {\bibinfo
  {volume} {103}},\ \bibinfo {pages} {124--130} (\bibinfo {year}
  {2015})}\BibitemShut {NoStop}%
\bibitem [{\citenamefont {Pearton}, \citenamefont {Abernathy},\ and\
  \citenamefont {Ren}(1994)}]{Pearton1994}%
  \BibitemOpen
  \bibfield  {author} {\bibinfo {author} {\bibfnamefont {S.~J.}\ \bibnamefont
  {Pearton}}, \bibinfo {author} {\bibfnamefont {C.~R.}\ \bibnamefont
  {Abernathy}}, \ and\ \bibinfo {author} {\bibfnamefont {F.}~\bibnamefont
  {Ren}},\ }\href@noop {} {\bibfield  {journal} {\bibinfo  {journal} {Applied
  Physics Letters}\ }\textbf {\bibinfo {volume} {64}} (\bibinfo {year}
  {1994})}\BibitemShut {NoStop}%
\bibitem [{\citenamefont {Sankaran}\ and\ \citenamefont
  {Giapis}(2001)}]{Sankaran2001}%
  \BibitemOpen
  \bibfield  {author} {\bibinfo {author} {\bibfnamefont {R.~M.}\ \bibnamefont
  {Sankaran}}\ and\ \bibinfo {author} {\bibfnamefont {K.~P.}\ \bibnamefont
  {Giapis}},\ }\href@noop {} {\bibfield  {journal} {\bibinfo  {journal}
  {Applied Physics Letters}\ }\textbf {\bibinfo {volume} {79}} (\bibinfo {year}
  {2001})}\BibitemShut {NoStop}%
\bibitem [{\citenamefont {Lukins}, \citenamefont {Zareie},\ and\ \citenamefont
  {Khachan}(2001)}]{Lukins2001}%
  \BibitemOpen
  \bibfield  {author} {\bibinfo {author} {\bibfnamefont {P.~B.}\ \bibnamefont
  {Lukins}}, \bibinfo {author} {\bibfnamefont {M.~H.}\ \bibnamefont {Zareie}},
  \ and\ \bibinfo {author} {\bibfnamefont {J.}~\bibnamefont {Khachan}},\
  }\href@noop {} {\bibfield  {journal} {\bibinfo  {journal} {Applied Physics
  Letters}\ }\textbf {\bibinfo {volume} {78}} (\bibinfo {year}
  {2001})}\BibitemShut {NoStop}%
\bibitem [{\citenamefont {Wu}\ and\ \citenamefont {Kumar}(2007)}]{Wu2007}%
  \BibitemOpen
  \bibfield  {author} {\bibinfo {author} {\bibfnamefont {B.}~\bibnamefont
  {Wu}}\ and\ \bibinfo {author} {\bibfnamefont {A.}~\bibnamefont {Kumar}},\
  }\href {\doibase http://dx.doi.org/10.1063/1.2470470} {\bibfield  {journal}
  {\bibinfo  {journal} {Applied Physics Letters}\ }\textbf {\bibinfo {volume}
  {90}},\ \bibinfo {eid} {063105} (\bibinfo {year} {2007})}\BibitemShut
  {NoStop}%
\bibitem [{\citenamefont {Kyung}(2007)}]{Kyung2007}%
  \BibitemOpen
  \bibfield  {author} {\bibinfo {author} {\bibfnamefont {P.~J.-B. L. J.-H. L.
  J.-T.}\ \bibnamefont {Kyung}, \bibfnamefont {Se-Jin}},\ }\href@noop {}
  {\bibfield  {journal} {\bibinfo  {journal} {Applied Physics Letters}\
  }\textbf {\bibinfo {volume} {91}},\ \bibinfo {pages} {091504 -- 091504--3}
  (\bibinfo {year} {2007})}\BibitemShut {NoStop}%
\bibitem [{\citenamefont {Deng}\ \emph {et~al.}(2014)\citenamefont {Deng},
  \citenamefont {Leys}, \citenamefont {Vujosevic}, \citenamefont {Vuksanovic},
  \citenamefont {Cvelbar}, \citenamefont {De~Geyter}, \citenamefont {Morent},\
  and\ \citenamefont {Nikiforov}}]{Deng2014}%
  \BibitemOpen
  \bibfield  {author} {\bibinfo {author} {\bibfnamefont {X.}~\bibnamefont
  {Deng}}, \bibinfo {author} {\bibfnamefont {C.}~\bibnamefont {Leys}}, \bibinfo
  {author} {\bibfnamefont {D.}~\bibnamefont {Vujosevic}}, \bibinfo {author}
  {\bibfnamefont {V.}~\bibnamefont {Vuksanovic}}, \bibinfo {author}
  {\bibfnamefont {U.}~\bibnamefont {Cvelbar}}, \bibinfo {author} {\bibfnamefont
  {N.}~\bibnamefont {De~Geyter}}, \bibinfo {author} {\bibfnamefont
  {R.}~\bibnamefont {Morent}}, \ and\ \bibinfo {author} {\bibfnamefont
  {A.}~\bibnamefont {Nikiforov}},\ }\href@noop {} {\bibfield  {journal}
  {\bibinfo  {journal} {Plasma Processes and Polymers}\ }\textbf {\bibinfo
  {volume} {11}},\ \bibinfo {pages} {921--930} (\bibinfo {year}
  {2014})}\BibitemShut {NoStop}%
\bibitem [{\citenamefont {Petrov}\ and\ \citenamefont
  {Zhechev}(2002)}]{Petrov2002}%
  \BibitemOpen
  \bibfield  {author} {\bibinfo {author} {\bibfnamefont {G.~M.}\ \bibnamefont
  {Petrov}}\ and\ \bibinfo {author} {\bibfnamefont {D.}~\bibnamefont
  {Zhechev}},\ }\href@noop {} {\bibfield  {journal} {\bibinfo  {journal}
  {Physics of Plasmas (1994-present)}\ }\textbf {\bibinfo {volume} {9}}
  (\bibinfo {year} {2002})}\BibitemShut {NoStop}%
\bibitem [{\citenamefont {Qi}\ \emph {et~al.}(2009)\citenamefont {Qi},
  \citenamefont {Huang}, \citenamefont {Gao},\ and\ \citenamefont
  {Qiu}}]{Qi2009}%
  \BibitemOpen
  \bibfield  {author} {\bibinfo {author} {\bibfnamefont {B.}~\bibnamefont
  {Qi}}, \bibinfo {author} {\bibfnamefont {J.}~\bibnamefont {Huang}}, \bibinfo
  {author} {\bibfnamefont {L.}~\bibnamefont {Gao}}, \ and\ \bibinfo {author}
  {\bibfnamefont {Y.}~\bibnamefont {Qiu}},\ }\href {\doibase
  http://dx.doi.org/10.1063/1.3200894} {\bibfield  {journal} {\bibinfo
  {journal} {Physics of Plasmas (1994-present)}\ }\textbf {\bibinfo {volume}
  {16}},\ \bibinfo {eid} {083301} (\bibinfo {year} {2009})}\BibitemShut
  {NoStop}%
\bibitem [{\citenamefont {Qi}\ \emph {et~al.}(2014)\citenamefont {Qi},
  \citenamefont {Pan}, \citenamefont {Zhou}, \citenamefont {Huang},\ and\
  \citenamefont {Liu}}]{Qi2014}%
  \BibitemOpen
  \bibfield  {author} {\bibinfo {author} {\bibfnamefont {B.}~\bibnamefont
  {Qi}}, \bibinfo {author} {\bibfnamefont {L.}~\bibnamefont {Pan}}, \bibinfo
  {author} {\bibfnamefont {Q.}~\bibnamefont {Zhou}}, \bibinfo {author}
  {\bibfnamefont {J.}~\bibnamefont {Huang}}, \ and\ \bibinfo {author}
  {\bibfnamefont {Y.}~\bibnamefont {Liu}},\ }\href {\doibase
  http://dx.doi.org/10.1063/1.4904377} {\bibfield  {journal} {\bibinfo
  {journal} {Physics of Plasmas (1994-present)}\ }\textbf {\bibinfo {volume}
  {21}},\ \bibinfo {eid} {123302} (\bibinfo {year} {2014})}\BibitemShut
  {NoStop}%
\bibitem [{\citenamefont {Faure}\ and\ \citenamefont
  {Shkol'nik}(1998)}]{Faure1998}%
  \BibitemOpen
  \bibfield  {author} {\bibinfo {author} {\bibfnamefont {G.}~\bibnamefont
  {Faure}}\ and\ \bibinfo {author} {\bibfnamefont {S.~M.}\ \bibnamefont
  {Shkol'nik}},\ }\href {http://stacks.iop.org/0022-3727/31/i=10/a=013}
  {\bibfield  {journal} {\bibinfo  {journal} {Journal of Physics D: Applied
  Physics}\ }\textbf {\bibinfo {volume} {31}},\ \bibinfo {pages} {1212}
  (\bibinfo {year} {1998})}\BibitemShut {NoStop}%
\bibitem [{\citenamefont {Cruden}\ \emph {et~al.}(2002)\citenamefont {Cruden},
  \citenamefont {Rao}, \citenamefont {Sharma},\ and\ \citenamefont
  {Meyyappan}}]{Cruden2002}%
  \BibitemOpen
  \bibfield  {author} {\bibinfo {author} {\bibfnamefont {B.~A.}\ \bibnamefont
  {Cruden}}, \bibinfo {author} {\bibfnamefont {M.~V. V.~S.}\ \bibnamefont
  {Rao}}, \bibinfo {author} {\bibfnamefont {S.~P.}\ \bibnamefont {Sharma}}, \
  and\ \bibinfo {author} {\bibfnamefont {M.}~\bibnamefont {Meyyappan}},\
  }\href@noop {} {\bibfield  {journal} {\bibinfo  {journal} {Applied Physics
  Letters}\ }\textbf {\bibinfo {volume} {81}} (\bibinfo {year}
  {2002})}\BibitemShut {NoStop}%
\bibitem [{\citenamefont {Staack}\ \emph {et~al.}(2005)\citenamefont {Staack},
  \citenamefont {Farouk}, \citenamefont {Gutsol},\ and\ \citenamefont
  {Fridman}}]{Staack2005}%
  \BibitemOpen
  \bibfield  {author} {\bibinfo {author} {\bibfnamefont {D.}~\bibnamefont
  {Staack}}, \bibinfo {author} {\bibfnamefont {B.}~\bibnamefont {Farouk}},
  \bibinfo {author} {\bibfnamefont {A.}~\bibnamefont {Gutsol}}, \ and\ \bibinfo
  {author} {\bibfnamefont {A.}~\bibnamefont {Fridman}},\ }\href
  {http://stacks.iop.org/0963-0252/14/i=4/a=009} {\bibfield  {journal}
  {\bibinfo  {journal} {Plasma Sources Science and Technology}\ }\textbf
  {\bibinfo {volume} {14}},\ \bibinfo {pages} {700} (\bibinfo {year}
  {2005})}\BibitemShut {NoStop}%
\bibitem [{\citenamefont {Bai}, \citenamefont {Sawin},\ and\ \citenamefont
  {Cruden}(2006)}]{Bai2006}%
  \BibitemOpen
  \bibfield  {author} {\bibinfo {author} {\bibfnamefont {B.}~\bibnamefont
  {Bai}}, \bibinfo {author} {\bibfnamefont {H.~H.}\ \bibnamefont {Sawin}}, \
  and\ \bibinfo {author} {\bibfnamefont {B.~A.}\ \bibnamefont {Cruden}},\
  }\href {\doibase 10.1063/1.2159545} {\bibfield  {journal} {\bibinfo
  {journal} {Journal of Applied Physics}\ }\textbf {\bibinfo {volume} {99}},\
  \bibinfo {eid} {013308} (\bibinfo {year} {2006})}\BibitemShut {NoStop}%
\bibitem [{\citenamefont {Lu}\ \emph {et~al.}(2013)\citenamefont {Lu},
  \citenamefont {Xu}, \citenamefont {Zhong}, \citenamefont {Ostrikov},
  \citenamefont {Cvelbar},\ and\ \citenamefont {Mariotti}}]{Lu2013}%
  \BibitemOpen
  \bibfield  {author} {\bibinfo {author} {\bibfnamefont {Y.}~\bibnamefont
  {Lu}}, \bibinfo {author} {\bibfnamefont {S.~F.}\ \bibnamefont {Xu}}, \bibinfo
  {author} {\bibfnamefont {X.~X.}\ \bibnamefont {Zhong}}, \bibinfo {author}
  {\bibfnamefont {K.}~\bibnamefont {Ostrikov}}, \bibinfo {author}
  {\bibfnamefont {U.}~\bibnamefont {Cvelbar}}, \ and\ \bibinfo {author}
  {\bibfnamefont {D.}~\bibnamefont {Mariotti}},\ }\href
  {http://stacks.iop.org/0295-5075/102/i=1/a=15002} {\bibfield  {journal}
  {\bibinfo  {journal} {EPL (Europhysics Letters)}\ }\textbf {\bibinfo {volume}
  {102}},\ \bibinfo {pages} {15002} (\bibinfo {year} {2013})}\BibitemShut
  {NoStop}%
\bibitem [{\citenamefont {Xu}\ and\ \citenamefont {Zhong}(2013)}]{Xu2013}%
  \BibitemOpen
  \bibfield  {author} {\bibinfo {author} {\bibfnamefont {S.~F.}\ \bibnamefont
  {Xu}}\ and\ \bibinfo {author} {\bibfnamefont {X.~X.}\ \bibnamefont {Zhong}},\
  }\href {\doibase http://dx.doi.org/10.1063/1.4813268} {\bibfield  {journal}
  {\bibinfo  {journal} {Applied Physics Letters}\ }\textbf {\bibinfo {volume}
  {103}},\ \bibinfo {eid} {024101} (\bibinfo {year} {2013})}\BibitemShut
  {NoStop}%
\bibitem [{\citenamefont {Zhou}\ \emph {et~al.}(2007)\citenamefont {Zhou},
  \citenamefont {Zhong}, \citenamefont {Wu}, \citenamefont {Yuan},
  \citenamefont {Shu}, \citenamefont {Li},\ and\ \citenamefont
  {Xia}}]{Zhou2007}%
  \BibitemOpen
  \bibfield  {author} {\bibinfo {author} {\bibfnamefont {W.}~\bibnamefont
  {Zhou}}, \bibinfo {author} {\bibfnamefont {X.}~\bibnamefont {Zhong}},
  \bibinfo {author} {\bibfnamefont {X.}~\bibnamefont {Wu}}, \bibinfo {author}
  {\bibfnamefont {L.}~\bibnamefont {Yuan}}, \bibinfo {author} {\bibfnamefont
  {Q.}~\bibnamefont {Shu}}, \bibinfo {author} {\bibfnamefont {W.}~\bibnamefont
  {Li}}, \ and\ \bibinfo {author} {\bibfnamefont {Y.}~\bibnamefont {Xia}},\
  }\href {http://stacks.iop.org/0022-3727/40/i=1/a=018} {\bibfield  {journal}
  {\bibinfo  {journal} {Journal of Physics D: Applied Physics}\ }\textbf
  {\bibinfo {volume} {40}},\ \bibinfo {pages} {219} (\bibinfo {year}
  {2007})}\BibitemShut {NoStop}%
\bibitem [{\citenamefont {Phillips}(1976)}]{Phillips1976}%
  \BibitemOpen
  \bibfield  {author} {\bibinfo {author} {\bibfnamefont {D.~M.}\ \bibnamefont
  {Phillips}},\ }\href {http://stacks.iop.org/0022-3727/9/i=3/a=017} {\bibfield
   {journal} {\bibinfo  {journal} {Journal of Physics D: Applied Physics}\
  }\textbf {\bibinfo {volume} {9}},\ \bibinfo {pages} {507} (\bibinfo {year}
  {1976})}\BibitemShut {NoStop}%
\bibitem [{\citenamefont {Hartmann}\ and\ \citenamefont
  {Johnson}(1978)}]{Hartmann1978}%
  \BibitemOpen
  \bibfield  {author} {\bibinfo {author} {\bibfnamefont {G.}~\bibnamefont
  {Hartmann}}\ and\ \bibinfo {author} {\bibfnamefont {P.~C.}\ \bibnamefont
  {Johnson}},\ }\href {http://stacks.iop.org/0022-3700/11/i=9/a=013} {\bibfield
   {journal} {\bibinfo  {journal} {Journal of Physics B: Atomic and Molecular
  Physics}\ }\textbf {\bibinfo {volume} {11}},\ \bibinfo {pages} {1597}
  (\bibinfo {year} {1978})}\BibitemShut {NoStop}%
\end{thebibliography}
\providecommand{\noopsort}[1]{}\providecommand{\singleletter}[1]{#1}%

\end{document}